\documentclass{article}

\usepackage[final]{neurips_2023_ml4ps}

\usepackage[utf8]{inputenc} 
\usepackage[T1]{fontenc}    
\usepackage{hyperref}       
\usepackage{url}            
\usepackage{booktabs}       
\usepackage{amsfonts}       
\usepackage{nicefrac}       
\usepackage{microtype}      
\usepackage{graphicx}
\usepackage{placeins}
\usepackage{multirow}
\usepackage{caption, subcaption}

\title{Autoregressive Transformers for Disruption Prediction in Nuclear Fusion Plasmas}

\author{%
  Lucas Spangher \\
  Plasma Science and Fusion Center \\
  Massachusetts Institute of Technology\\
  \texttt{spangher@psfc.mit.edu} \\
  \And
  William F. Arnold \\
  Kim Jaechul School of AI \\
  KAIST \\
  \texttt{will@mli.kaist.ac.kr} \\
  \And
  Alexander Spangher \\
  Information Sciences Institute \\
  University of Southern California 
  \And
  Andrew Maris \\
  Plasma Science and Fusion Center \\
  Massachusetts Institute of Technology\\
  \And 
  Cristina Rea\\
  Plasma Science and Fusion Center \\
  Massachusetts Institute of Technology\\
  \texttt{crea@psfc.mit.edu}
}

\begin{document}

\maketitle

\begin{abstract}


The physical sciences require models tailored to specific nuances of different dynamics. In this work, we study outcome predictions in nuclear fusion tokamaks, where a major challenge are \textit{disruptions}, or the loss of plasma stability with damaging implications for the tokamak. Although disruptions are difficult to model using physical simulations, machine learning (ML) models have shown promise in predicting these phenomena. Here, we first study several variations on masked autoregressive transformers, achieving an average of 5\% increase in Area Under the Receiving Operating Characteristic metric above existing methods. We then compare transformer models to limited context neural networks in order to shed light on the ``memory'' of plasma effected by tokamaks controls. With these model comparisons, we argue for the persistence of a memory throughout the plasma \textit{in the context of tokamaks} that our model exploits. 

\end{abstract}

\section{Introduction}

Nuclear fusion has been long sought as a promising method of generating carbon-free energy without major land-use requirements or high-level radioactive waste. One major approach to obtaining fusion in laboratory plasmas is through \textit{magnetic confinement}, which leverages the tendency of plasmas to remain confined perpendicular to the magnetic field lines. The fusion approach with the highest technological readiness level \cite{Wurzel2021ProgressTF} is the \textit{tokamak}, a toroidal chamber with external magnetic coils and a large current that runs toroidally through the plasma with the goal of generating net positive energy (see Figure \ref{fig:seq2lab}.)

A major problem facing fusion tokamaks are \textit{disruptions}, or loss of plasma stability with damaging implications for the tokamak. Disruptions are difficult to model using physical simulations due in part to the heterogeneity of causes that can lead to instability, the multi-scale nature of plasma dynamics, and to a wealth of unobserved factors that affect known physical dynamics. Thus, there is an emerging literature on methods other than ``first principle models'' for disruption prediction.

The goal in this prediction problem is to provide sufficient time to engage control actions or emergency  mitigation systems, which curb the deleterious consequences of disruptions.  Numerous ML models have been proposed to predict the likelihood that a plasma disruption will occur \cite{cannas2013automatic, kates2019predicting}. Prior work has investigated the use of random forest models to monitor plasma states \cite{reaRealtimeMachineLearningbased2019} atlargest tokamak in the US, DIII-D, followed by current state of the art predictors, such as the Hybrid Deep Learner (HDL) proposed in 2021, which utilize classical neural techniques such as a convolutional LSTM network \cite{zhuHybridDeeplearningArchitecture2020}. 

However, none of these models have the desired accuracy for commercial scale tokamaks where disruption prediction accuracy must be nearly flawless (>95\% true positive rate) \cite{maris2023impact}. Also, we hypothesize, none are tuned to the specific needs of the problem. It is unclear how long the ``memory'' of the plasma is -- while some argue that as little as 200 milliseconds of history is needed to capture the evolution of the plasma temperature and density \cite{zhuHybridDeeplearningArchitecture2020, reaRealtimeMachineLearningbased2019}, others argue that certain instabilities or tokamak control errors that occur near the start of a seconds-long shot may impact the end behavior of the shots\cite{cannas2013automatic}. 

In this work, we apply variations on the Transformer architecture \cite{vaswani2017attention, bengio2009curriculum}, which has become popular in recent years to model long term, event-based time-series. Our two contributions are:

\begin{enumerate}
\item We create a state-of-the-art model on a data from three major tokamaks. We find an average of .05 improvement in Area under the Receiver Operating Characteristic Curve (AUC) over the HDL baseline (Table \ref{tab:AUCs}.) We find various pretraining steps help augment performance.
\item We investigate whether long term memory is helpful in modeling the fusion reaction, giving us scientific insight into the dynamics of the process. 
\end{enumerate}

\section{Dataset Provenance and Preparation}
\label{sct:dataset}

Our dataset is composed of trials from three different tokamaks, $r_1$, $r_2$ and $r_3$.\footnote{$r_1$ is MIT's Alcator C-Mod, $r_2$ is General Atomic's DIII-D, and $r_3$ is the Chinese Academy of Science's EAST tokamak}. Each $r_i$ differs considerably\footnote{For a longer discussion on the physical differences between the three, see \cite{montes2019machine}}.
Each tokamak is run continuously for a ``shot'' sequence lasting seconds to minutes, generating gigabytes of multi-modal data, of which we focus on global state variables (the rest include 1D states ("profiles"), 2D cross-sections, and camera photos); we focus on shots' global state variables, listed in Table \ref{tab:datasets}. Our objective is to understand how well this model would deploy to a new tokamak, which is 
similar to $r_1$. Thus, we restrict our test data to data generated by $r_1$. We build off previous work from our lab for the training set composition, which contains different proportions of $r_1$, $r_2$, and $r_3$ data [citation withheld]. 

Each shot is of length $n$, and we discard shots that are shorter than 125ms. Each tokamak's diagnostics store metrics from each shot at a different sampling rate. We normalize across tokamaks by discretizing at 5ms and interpolating via forward-fill. Roughly 10\% of shots end in disruptions. We truncate our time series to $X=1,...,n-\nu$, where $n$ is the full length and $\nu=40ms$, the minimum amount of time for a disruption mitigation system to activate. For a description of the features chosen, please see Table \ref{tab:features} in the Appendix.

\section{Model Details}

\begin{figure}
    \centering
    \includegraphics[width=\linewidth]{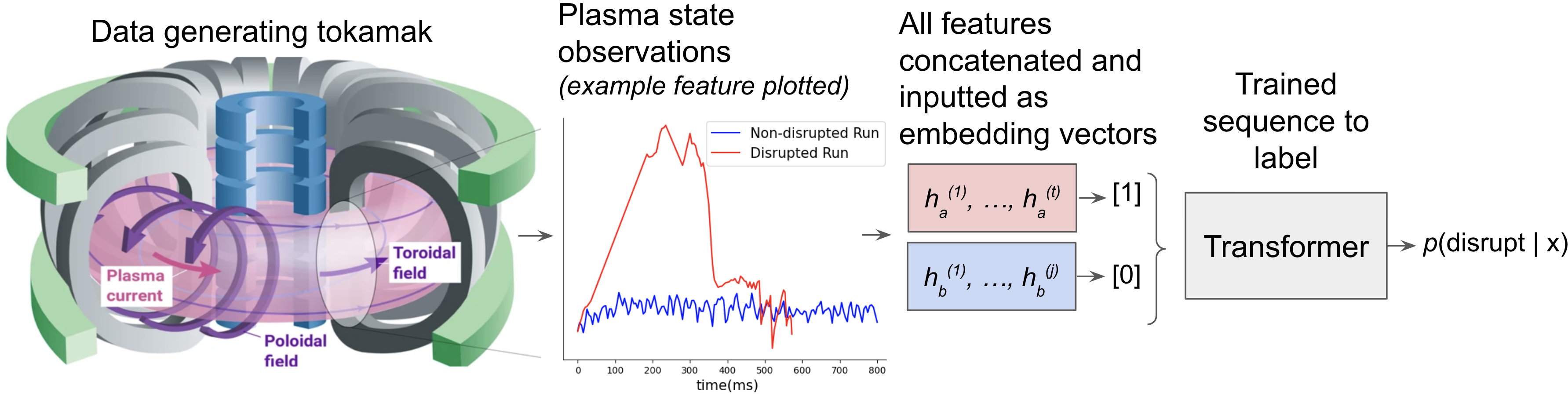}
    \caption{Sequence to Label classification task using GPT2-like transformers.}
    \label{fig:seq2lab}
\end{figure}

\subsection{Modeling Task}

We seek to perform label prediction (Figure \ref{fig:seq2lab}), where $X$ is described above, and $y$ is 1 if the shot ended in disruption and 0 otherwise. 
We follow \cite{zhuHybridDeeplearningArchitecture2020} in several augmentations. First, we generate additional pairs $(\overline{X}, \overline{y})$ from each $(X, y)$ by randomly sampled starting and ending windows of the shot. Next, in order enforce greater reasoning about the start of instabilities, we set a $\tau_{r_i}$ variable per tokamak\footnote{$\tau(r_i)$ is  the average length of time before a disruption occurs which is tokamak-dependent, see Table \ref{tab:datasets}}: if $(\overline{X}, \overline{y})$ of length $n$ ends before $n - \tau_{r_i}$, it is labelled 0.

\subsection{Model Architecture}

We adapt an autoregressive transformer based on  GPT\footnote{We utilize Huggingface's implementation: \url{https://github.com/huggingface/transformers/blob/main/src/transformers/models/gpt2/modeling_gpt2.py}} \cite{radford2018improving}. GPT-based models, which are typically used to model long sequences of discrete word-tokens, have an embedding layer that is used to project discrete tokens into a continuous embedding space. The dimensions of our data are already continuous variables, so we treat them as embeddings directly. These vectors are normalized, based on training-set mean and variance.

We perform a hyperparameter sweep using a bayesian hyperop strategy \cite{wu2019hyperparameter} over number of layers, dropout proportions, learning rate schedulers, optimizer $\beta$ parameters, and other hyperparameters of interest. All training was done on a cluster of 48 A100 Nvidia GPUs and tested on Area Under Receiver Operator Characteristc (AUC). 

\subsection{Model Augmentations: State Pretraining and Data-cutoff Currica Induction}

\textbf{State Pretraining}: We pretrain the model first on a next state prediction (NSP) task. We model NSP with a regression head on the model and Mean Average Error (MAE) loss. After convergence, which is observed to be 50k steps, we replace the regression head with a classification head and perform our sequence classification task. 

\textbf{Data-cutoff Curriculum Learning}: We notice the model often misclassifies samples with rapidly evolving end dynamics. We theorize that exposing the model to the full data, without the 40ms cutoff, might improve its predictions for the final 40ms. Using \textit{curriculum learning} \cite{bengio2009curriculum}, we sequentially train the model on easier tasks: predicting datasets with cutoffs from 0ms to 30ms, in 10ms increments, for 50k steps each. We then test the model on the main task. For example, a 3 step curriculum trains the model on data with 20ms cutoff, then 30ms, then 40.  We experiment with varying curriculum lengths from zero to five steps.

\section{Results and Discussion}

\subsection{Performance metrics}

Compared to the baseline HDL predictor, our vanilla model achieves favorable AUC across all cases, and our augmented model improves further on the vanilla case (Table \ref{tab:AUCs}.) Wall-clock time on forward inference ranges from 1-3ms, which is suitable for real-time deployment. 

\begin{table}[t]
    \centering
    \renewcommand{\arraystretch}{1.2} 
    \begin{tabular}{cccccccc}
    \toprule
    & \multicolumn{4}{c}{Training Set Composition} & \multicolumn{3}{c}{AUCs} \\
    \cmidrule(lr){2-5} \cmidrule(lr){6-8}
    & \multicolumn{2}{c}{Existing: $r^2, r^3$} & \multicolumn{2}{c}{ ``New'': $r^1$} & HDL & GPT & Aug-GPT \\
    \cmidrule(lr){2-3} \cmidrule(lr){4-5} \cmidrule(lr){6-6} \cmidrule(lr){7-7} \cmidrule(lr){8-8}
    Case & Non-disruptions (ND) & Disruptions (D) & ND & D &  & & \\
    \midrule
    1 & None & None & All & 20 & .640 & .781 & \textbf{.790} \\
    2 & None & None & All & All & .799 & .836 & \textbf{.843} \\ 
    3 & None & All & All & All & .808 & .831 & \textbf{.841} \\ 
    4 & All & All & None & None & .588 & \textbf{.622} &.621 \\ 
    5 & All & All & All & 20 & .703 & .699 & \textbf{.704} \\
    6 & All & All & All & All & .794 & .803 & \textbf{.814} \\\hline
    \addlinespace[0.5ex] 
    Mean & & & & & .72 & .76 & \textbf{.77} \\
    \bottomrule
    \end{tabular}
    \caption{A table containing different training cases and output metrics between our transformer and HDL, the baseline. Ours improves prior SOTA by 5\% AUC on average. Please note that ``GPT'' refers to the vanilla GPT, and ``Aug-GPT'' refers to the augmented GPT.}
    \label{tab:AUCs}
\end{table}

\subsection{Curriculum Training and State Pretraining}

\textbf{Curriculum Learning}: We observe, interestingly, that the model achieves the highest accuracy, recall, and precision ($\sim 10\%$ gain in each) following a curriculum in which the models start training 10ms from shots' ends and progress successively -- i.e. four curriculum steps. The result matches physical intuition: at the very end of disruptive shots, the plasma has already touched the tokamak's walls, and thus the physics of the fluid have changed. 

\textbf{State pretraining}: we observe that model performance consistently improves. We also observe that the transformer model seems to be a very good state predictor, which may be a side benefit of the model for other uses. 

\subsection{Long Term Memory}

\begin{figure}[t]
    \includegraphics[width=\linewidth]{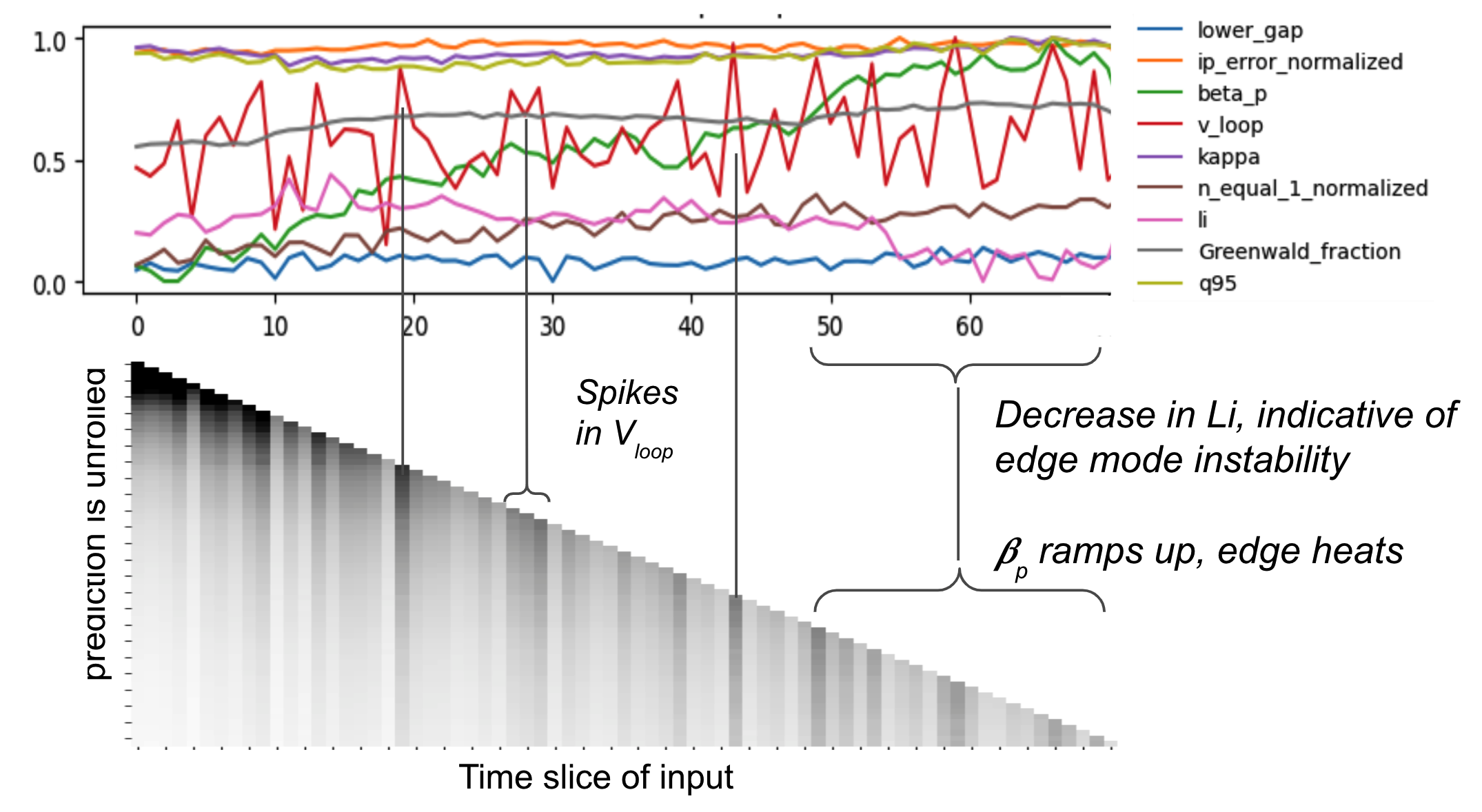}
    \label{fig:attention-map}

    \caption{Here we show a further exploration of transformer behavior. In the top plot, we show normalized input features, corresponding to the transformer's attention map below. To read the attention map, an unrolling sequence starts at the top left and then progresses by traversing down.}
    \label{fig:attention}
\end{figure}

We argue that plasma time-series exhibit long-term memory through three ways: 

\begin{enumerate}
    \item  A \textbf{comparison of model performance} finds the vanilla GPT significantly outperforming the baseline HDL, which explicitly encodes limited memory. 
    \item \textbf{Ablations on data context} show that model performance decreases as available context decreases from 480ms to 80ms, indicating that the model may be using data context from throughout the shot (Figure \ref{fig:attention}.) 
    \item \textbf{Global attention maps}, which explicitly encode regions of the data to which the model pays attention, does not tend towards zero even when the whole shot is observed (Figure \ref{fig:attention})
\end{enumerate}

\subsection{Model interpretation.}

Generally we may interpret characteristics of the data that the transformer picks out. First, we find the attention spikes maps to loop voltage spikes, of which higher voltage generally correlates to increased system pressure. Second, we find that shaded regions of the attention map correspond to an uptick in $\beta_p$ and a drop in $Li$. These characteristics are consistent with those of a cooling plasma prone for edge instabilities \cite{zohm1996edge}. Both of these observations give confidence that the transformer is picking up on known physical events. 

\begin{figure}[!htb]
    \centering
    \includegraphics[width=.6\linewidth]{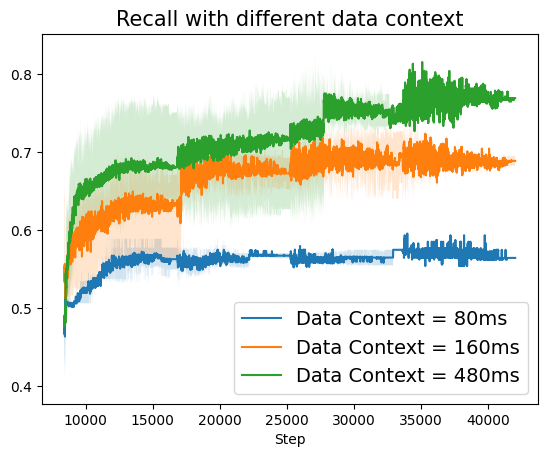}
    \caption{Exploration of transformer behavior. Here, we ablate on limiting data context to test whether performance is directly effected by physical memory.}
    \label{fig:ctx-plot}
\end{figure}

\section{Conclusion and Future Work}

Transformers are promising models for the physical sciences. We present a case where of transformers, as well as augmentations, achieve better performance than baseline. We demonstrate that the model picks up on some physics, and benefits from explicit memory of the system. 

One shortcoming of the model is that it is optimized for discrete time series, like sentences, where words are some fundamental unit of the process. Ours, meanwhile, is a time series characterizing a dynamic process, where sampling rate is relatively arbitrary and may change, not only between tokamaks, but between diagnostics within a single tokamak. While impressive that the transformer can learn attention on discrete spikes in $v_{loop}$, we ultimately desire a model that may capture continuous patterns and subtle variations. Thus in future work, we hope to evaluate the performance of an Autoformer \cite{wu2022autoformer}, which is a transformer whose discrete attention blocks are replaced with autocorrelational FFT decompositions, making it more suited for continuous sequences.

\begin{ack}
We thank PSFC for providing ample compute resources and Matteo Bonotto, Tommasso Gallingani, Daniele Bigone, and Francesco Cannarile for consistent technical collaboration, advice, ideas and feedback. This work was supported by Eni S.p.A. through the MIT Energy Initiative and by Commonwealth Fusion Systems under SPARC RPP021 funding. We are grateful to Robert Granetz for advice on physical interpretation of the model. 
\end{ack}
\bibliographystyle{plainnat}
\bibliography{ref}

\newpage
\FloatBarrier
\appendix

\section{Data Feature Decsription}

\begin{table}[!htb]
    \centering
    \begin{tabular}{|c|c|c|}\hline
        Feature & Definition & Relevant instab. \\ \hline \hline

           Locked mode indicator& Locked mode mag. field normalized to toroidal field & MHD \\  

          Rotating mode indicator & Std. dev. of Mirnov array normalized by toroidal field & MHD \\ 

         $\beta_p$ & Plasma pressure normalized by poloidal magnetic field & MHD \\

         $\ell_i$ & Normalized plasma internal inductance & MHD \\ 
         
         $q_{95}$ & Safety factor at 95th flux surface & MHD \\ \hline

         $n/n_G$ & Electron density normalized by Greenwald density & Dens. limit \\ 

         $\Delta z_{\textrm{center}}$ & Vertical position error of plasma current centroid & Vert. Stab. \\ 

         $\Delta z_{\textrm{lower}}$ & Gap between plasma and lower divertor & Shaping \\  

         $\kappa$ & Plasma elongation & Shaping \\ \hline

         $P_{\textrm{rad}}/P_{\textrm{input}}$ & Radiated power normalized by input power & Impurities \\ 

        $I_{\textrm{p,error}}/I_{\textrm{p,prog}} $ & Plasma current error normalized by programmed current & Impurities \\ 

         $V_{\textrm{loop}}$ & Toroidal ``loop'' voltage & Impurities \\ \hline

    \end{tabular}
    \caption{The input features of the model, their definitions, and a categorization of the type of instability the signal indicates.}
    \label{tab:features}
\end{table}

\section{Further Experimental Details}

\begin{table}[!ht]
    \centering
    \begin{tabular}{| c|c | c| c |c  |}\hline
        tokamak & $\tau$ & Number of Shots & Average Shot Length & Initial Sampling Rate \\\hline
        C-Mod &  50 ms & 4000 & .52s & .005ms \\
         DIII-D & 150 ms & 8000 & 3.7s& .01ms \\
         EAST & 400 ms & 11000 & 5.3s & .025ms \\\hline
    \end{tabular}
    \caption{Metrics on a dataset composed of multiple tokamaks.}
    \label{tab:datasets}
\end{table}



\section{Supplemental Background on Fusion Devices and Plasma Dynamics}

The two major approaches to generation are \textit{inertial confinement}, which uses lasers or motion to heat up plasma\cite{edwards2013progress}, and \textit{magnetic confinement}, which leverages extreme magnetic fields to pressurize gas into plasma, and then pressurize plasma into ``burning plasma'', i.e. ones that undergo nuclear fusion \cite{creely2020overview}.

Tokamaks exert magnetic fields of roughly 5 Tesla in a toroidal vacuum chamber filled with a gas of hydrogren, deuterium and tritium isoptopes\footnote{Deuterium and tritium isotopes are hydrogen atoms that have one and two extra neutrons, respectively}. Under extreme conditions, this gas enters a plasma state, which is characterized by the highly charged flow of subatomic particles (ions and electrons.) Tokamaks achieve some degree of magnetic-field symmetry, so tokamak plasmas are stable under certain conditions. However, magnetohydrodynamic (MHD) instabilities may still arise. For example, a majority of electrons may slow in toroidal rotations relative to poloidal rotations, creating areas where the induced magnetic force of the plasma pressures certain points of the toroid and not others. Left untreated, instabilities like this may grow out of control and lead to a loss of plasma confinement, causing an end to the reaction and a fast quench of the temperature and current onto the walls of the tokamak. Given that the generated plasma has a core temperature on par with the center of the sun, this may be a serious problem, resulting in damage to the walls and downtime, making plasma disruptions a fundamental hurdle facing tokamak commercializability \cite{spangher2019characterizing}.

\end{document}